\documentclass[prl,aps,twocolumn]{revtex4}
\usepackage{epsfig}
\usepackage{citesort}

\newcommand{\beq}{\begin{equation}} \newcommand{\eeq}{\end{equation}} 
\newcommand{\beqas}{\begin{eqnarray*}} \newcommand{\eeqas}{\end{eqnarray*}}

\begin{document} 
 
\title{ 
Hydrodynamic theory for dissipative hard spheres  
                with multi-particle interactions 
} 
 
\author{ Stefan Luding (1,2) and Alexander Goldshtein (3) } 
\address{ (1) Institute for Computer Applications 1,  
              Pfaffenwaldring 27, 70569 Stuttgart, GERMANY\\ 
          (2) Particle Technology, DelftChemTech, TUDelft,
              Julianalaan 136, 2628 BL Delft, NETHERLANDS\\
          (3) Faculty of Mechanical Engineering, Technion, 
              Haifa 32000, ISRAEL} 
 
 
\begin{abstract} 
Extension to kinetic theory and hydrodynamic models are proposed 
that account for the existence of multi-particle contacts.  
In the presence of multi-particle contacts (involving elastic, reversible,
potential contact energy), dissipation of the translational (kinetic) energy 
is reduced and a class of different models lead to deviations from 
the classical inelastic hard sphere (IHS) homogeneous cooling state (HCS),
as examined here. The theoretical results are found to be in perfect 
agreement with the numerical simulations.  
\end{abstract} 
 

\maketitle 
 
Kinetic theory and the related hydrodynamic models are helpful tools for the 
modeling and understanding of transport processes in classical, elastic gases 
for low and moderate densities \cite{chapman60,hansen86}. The hard sphere 
(HS) model is the corresponding approach to be implemented as a numerical 
model \cite{lowen00,luding02}.   
A successful theoretical approach in the spirit of Boltzmann 
or Chapman and Enskog \cite{chapman60,hansen86} requires the basic 
assumptions:  
(i) The collisions are instantaneous and  
(ii) subsequent collisions are uncorrelated (``molecular chaos'').   
Conditions (i) and (ii) lead to the Boltzmann equation, and in the 
equilibrium state, the velocity distribution is (iii) a Maxwellian.

When dissipation is added, one has the inelastic hard sphere (IHS) 
model, where the coefficient of restitution
$r$ quantifies dissipation, elastic systems have $r=1$, and $1-r^2>0$
determines the amount of energy lost in a two-particle collision
in the center of mass system.
The range of applicability of the theory for the IHS was addressed
in several papers \cite{goldshtein95,brey9697,noije98c,brilliantov01}.
However, this is far from the scope of this study, since it involves 
e.g.\ Sonine polynomial expansions and ``viscoelastic'' material
behavior \cite{brilliantov01}, so that we just assume (i--iii) 
valid for the sake of simplicity.
 
In this study we restrict ourselves to the homogeneous cooling  
state (HCS) and focus on a mean-field hydrodynamic approach 
\cite{goldshtein95,brey9697,sela98,noije98c,huthmann98,luding98d},
neglecting spatial structures like clusters or shear modes. 
This idealization is reasonable for either weak dissipation, 
low density or small system size, however, we do not discuss 
the range of applicability here.
The qualitative prediction for the long-time decay of energy 
with time by Haff \cite{haff83} is confirmed and
it was shown that the distribution function is isotropic in  
velocity space and it is close to a Maxwellian as long as 
the system is homogeneous \cite{huthmann00,brilliantov01,nie02}. 
 
\begin{figure}[htb]
\begin{center}
\epsfig{file=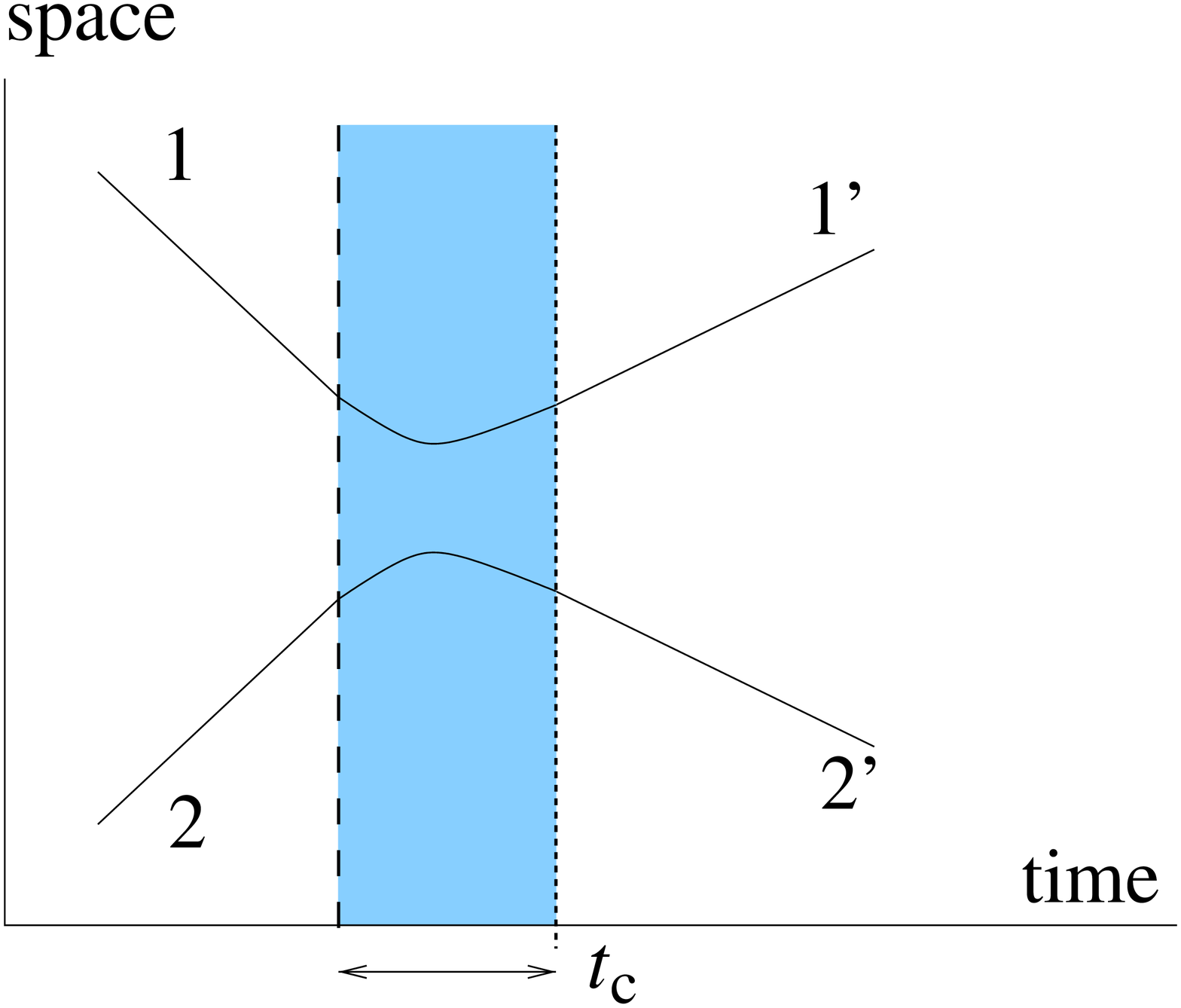,height=3.3cm,angle=0}
\epsfig{file=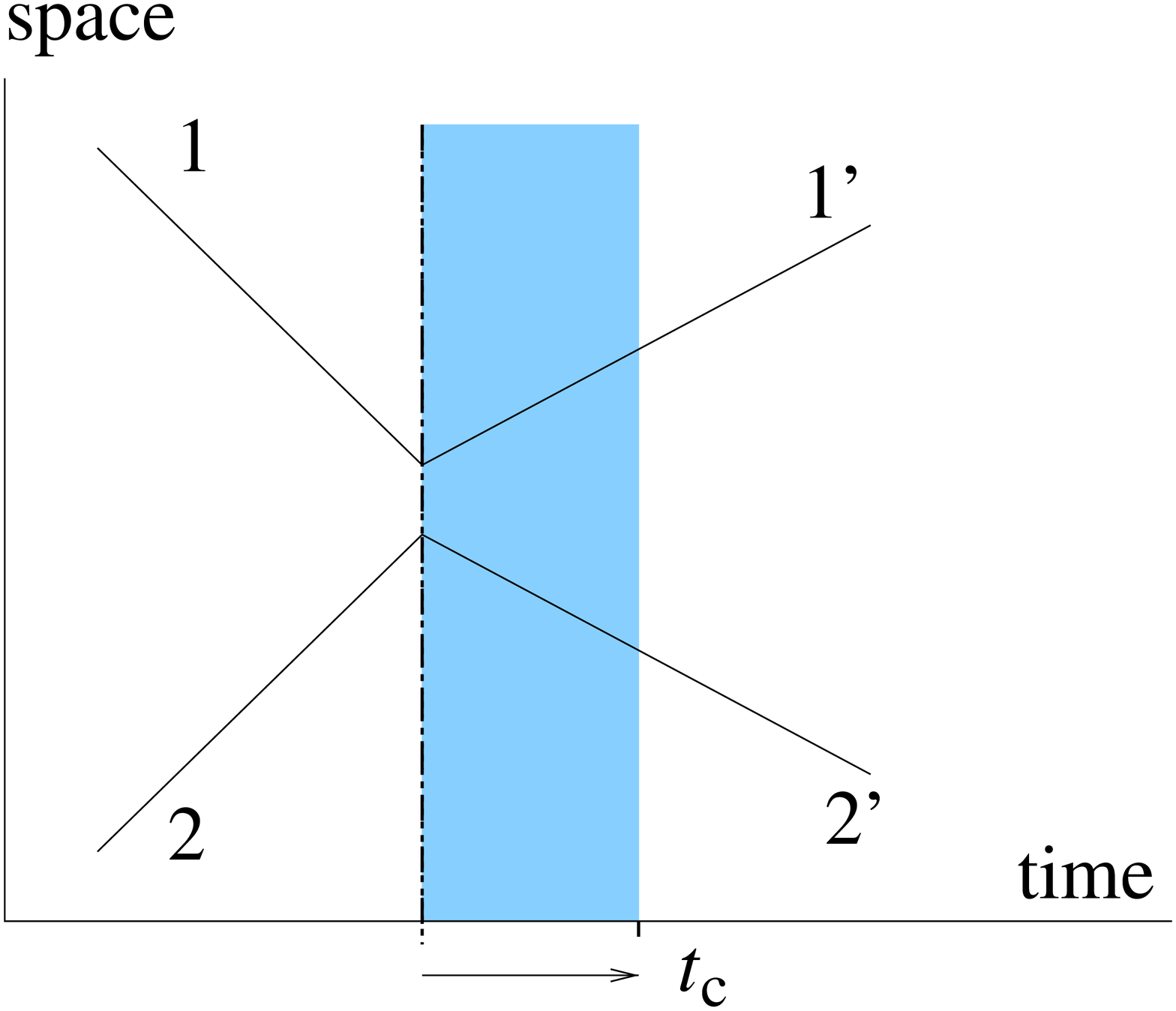,height=3.3cm,angle=0}
\end{center}
\caption{Schematic plot of the trajectories of two soft (left) and
two hard (right) particles against time. The beginning and the ending
of the interaction are marked by dashed and dotted vertical lines,
respectively, and the time $t_c$ during that kinetic energy is 
stored as elastic energy in the contact - so that dissipation is 
affected - is marked as shaded
region.
}
\label{fig:schem1}
\end{figure}

In the IHS collisions are always instantaneous,
see condition (i), due to the rigid interaction potential.  On the first
glance, this makes the model (and kinetic theory too) inadequate for
the description of real materials for which the interaction potential
may be steep, but is {\em never} perfectly rigid,
see Fig.\ \ref{fig:schem1}.
During the contact of two real particles, kinetic energy is stored
in elastic (reversible) potential energy that, in the static limit,
can be recovered after very long times. The conclusion is
thus that one has a fraction of elastic energy, {\em which cannot be 
dissipated}, in the real system. This fraction is missing in all 
idealized models HS, IHS, and also in the kinetic theory, and 
has to be defined.
Thus we will propose and examine possible ways to cure this 
problem of the hard sphere model, but keeping kinetic theory 
still applicable. 

The first step is to define or identify possible multi-particle 
contacts.  In a real system (or in a soft-particle model) one just 
counts the number of contacts a particle has. 
Within the extented IHS model, a particle remembers its last contact and
every contact occuring within a time $t_c$ after that, see the shaded
area in Fig.\ \ref{fig:schem1}, is defined a multi-particle contact.

In low density systems, where the mean free flight time is much larger 
than the contact duration, multi-particle contacts are rare.
However, assumption (i) can also be valid in high density situations 
where the free path is much smaller than the particle diameter: 
This is possible in the case of an extremely short contact duration, when
collisions remain practically instantaneous. Thus we conclude that
the free path, i.e.\ the density, is not an appropriate measure for
the occurence of multi-particle contacts.
We rather define, as a more objective criterion, the ratio  
$ \tau_c = {t_c}/{t_E} $
between the contact duration $t_c$ and the typical time between 
collisions $t_E$ as obtained by the Enskog theory 
\cite{chapman60,noije98c,luding98}.  Small and large $\tau_c$ values 
correspond to pair- and multi-particle collisions, respectively,
in our framework
\cite{luding98}.  
 
Up to now, the elastic HS model is {\em not} changed at all concerning
particle trajectories or whatsoever. The only modification is, that 
every particle that had a collision a short time ago keeps this event
in memory for the contact duration $t_c$ and every new contact occuring
within this time is now defined as an elastic contact.  This allows 
to split the total energy in the system into a
kinetic and an elastic (potential) contact energy \cite{luding98f},
as can be done in a real system. If a part of the kinetic 
energy is transfered to contact energy, it cannot be dissipated 
anymore, so that energy dissipation in the IHS model has to be
reduced in the presence of multi-particle contacts.  This qualitative
reduction of energy dissipation in dense systems has been 
observed in soft-particle molecular dynamics simulations \cite{luding94d}.
 
Recently, the transition of a granular gas to its solid  
counterpart has been investigated  \cite{luding98f,goldshtein03}.
A general model was defined, in which the coefficient of the normal 
restitution is given by 
\begin{equation} 
 r = r(x) =  
    \left \lbrace { 
      \begin{array}{ll} 
        r_0 ~, &{\rm ~~~if~~}  x >   x_c \\ 
        1   ~, &{\rm ~~~if~~}  x \le x_c 
      \end{array} 
     } \right . 
\label{eq:r0}
\end{equation} 
where $x$ is a variable and $x_c$ is the respective cut-off value. 
Above the threshold, one has the usual inelastic hard sphere mode 
with constant restitution coefficient, below the cut-off an elastic 
system with $r=1$ is assumed.   
The variables proposed were either the time between collisions (TC model),  
the distance travelled since the last collision (LC model), or the relative  
velocity of two particles prior to a collision (VC model).   
In order to keep the following  
analysis simple, we focus on the special case of a piecewise constant 
restitution and disregard any continuous dependency of $r$ on $x$ 
\footnote{ 
This simplification also is reasonable in the spirit and the framework 
of the kinetic theory and numerical event-driven simulations, where changes 
of the particle velocities occur as instantaneous, discontinuous events}. 
 
Variants of the general model have been used
\cite{mcnamara96,luding98f,ben-naim99,luding99,kamenetsky00,goldshtein02,nie02},
mainly to avoid the ``inelastic collapse'', an artefact of the rigid 
particle model, which allows an infinite number of collisions to occur 
within a finite time.  In the real system this can never occur due to the 
fact that the contacts are finite.
The model in Eq.\ (\ref{eq:r0}), avoids the inelastic collapse,
since the dense parts of the system, where the collapse tends  
to occur first, are transformed into elastic regions where  
the inelastic collapse is unlikely.   Thus the inelastic collapse
is replaced by static, dense regions of the material.
 
The different variants of the cut-off model will be discussed  
separately in the following, because they lead to different 
forms of the collision integral.  The general form of the  
energy balance equation is 
\begin{equation} 
\frac{d}{d \tau} E = -2 I(E,x_c) ~, 
\label{eq:dEdt} 
\end{equation} 
with the dimensionless time $\tau=(2/3)At/t_E(0)$,
scaled by $A=(1-r^2)/4$, and the initial collision rate $t_E(0)$.
In these units, the energy dissipation rate $I$ is a function of the  
dimensionless energy $E=K/K(0)$ with the kinetic energy $K$, and the 
cut-off parameter $x_c$.  In this representation, the restitution
coefficient is hidden in the rescaled time via $A$, so that IHS
simulations with different $r$ scale on the master-curve in the following 
plots.  In the following, we will extract the classical dissipation 
rate $E^{3/2}$ \cite{haff83} from $I$, so that  
\begin{equation} 
I(E,x_c) = J(E,x_c) E^{3/2}~, 
\label{eq:Jdef} 
\end{equation} 
with the correction-function $J \rightarrow 1$ for $x_c \rightarrow 0$. 
Our theoretical results will be compared with numerical  
simulations and with previous results \cite{luding98f}. 
For the derivation of the dimensionless equation (\ref{eq:dEdt})
from the kinetic theory in its dimensional form, 
see Refs.\ \cite{goldshtein02,luding98f}.

For the classical IHS model in the HCS, 
Eq.\ (\ref{eq:dEdt}) is solved by
$E_\tau=(1+\tau)^{-2}$, 
a master curve, independent of the coefficient 
of restitution $r$ and all other system parameters.
We checked via simulations 
that different $r$ values scale on the same master-curve,
as long as no clustering is obtained. 
We will proceed to develop our theory in the dimensionless
variables and will examine in detail the deviations from
the classical HCS.\\

The {\em velocity cut-off (VC) model} can be rationalized based on the 
picture of elasto-plastic particles which do not suffer inelastic (plastic) 
deformation if they collide below a certain threshold velocity $v_c$. 
(In static contact, the relative velocity vanishes and thus is automatically 
smaller than $v_c$), see \cite{nie02} for a recent application.
The deviation from the classical inelastic hard sphere HCS,
\begin{equation} 
J(E,v_c) = \left ( 1+\xi^2 \right ) 
           \exp \left ( -\xi^2 \right )~, 
\label{eq:Jvc} 
\end{equation} 
is obtained from the (cumbersome) computation of the collision integral
\cite{goldshtein03}, with the nondimensional quantity
\begin{equation} 
\xi^2 = \frac{3 m v_c^2}{8 K(t)}  
      = \left ( \frac{v_c^2}{4 v_T^2} \right )
      = \frac{V_c^2}{E} ~ \propto E^{-1} ~, 
\label{eq:defxi}
\end{equation} 
which relates the critical velocity to the actual mean  
fluctuation velocity.  The dimensionless cut-off velocity 
is $V_c={v_c/2v_T(0)}$.
For $v_c=0$ and thus $\xi=0$, the classical homogeneous cooling  
state is recovered, i.e.~$J(E,0)=1$. 
 
\begin{figure}[htb] 
\includegraphics[width=5.8cm,angle=-90]{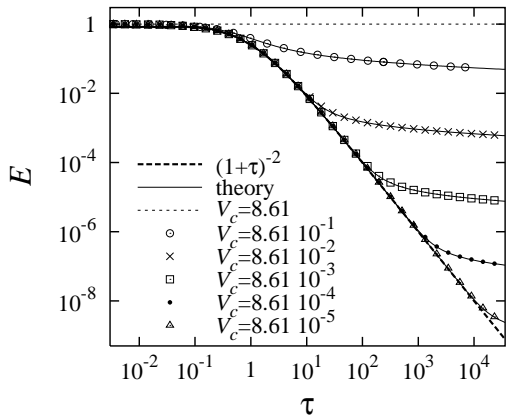} 
~\vspace{0.2cm}\\
\caption{Energy $E$ plotted against time $\tau$ for a simulation 
with $N=19683$ particles, density $\nu=0.08245$, coefficient 
of restitution $r_0=0.99$ and different $V_c$ as given in 
the inset. The symbols are simulations and the lines give the 
solution of Eqs.\ (\ref{eq:dEdt},\ref{eq:Jdef}), using 
Eqs.\ (\ref{eq:Jvc},\ref{eq:defxi}). 
} 
\label{fig:vc1} 
\end{figure}

Event driven numerical simulations \cite{luding98d,luding98f} 
are compared to the numerical solution of our theory in Fig.\ \ref{fig:vc1}. 
We obtain perfect agreement between theory and simulations  
in the examined range of $v_c$-values. The fixed cut-off velocity 
has no effect when the collision velocities are very large, $v_T \gg v_c$, 
but strongly reduces dissipation when the relative velocity at collision 
is comparable to or smaller than $v_c$.  Thus, in the homogeneous cooling 
state, there is no effect initially, but the long time behavior 
changes from the classical decay $E \propto t^{-2}$ to a constant,
time independent $E \propto t^0$, with extremely slow approach to
the constant. 
 
The {\em free path cut-off (LC) model} was used first by McNamara and Young 
\cite{mcnamara96} to avoid the inelastic collapse, and was extended  
to its simple hydrodynamic analogon expressed in terms of the density 
by Kamenetsky et al. \cite{kamenetsky00,goldshtein02} to describe the gas 
solid transition caused by the compression of a granular gas. 
The physical idea is that particles thath are near to each other 
-- their distance is below a certain threshold which can be  
regarded as some surface roughness -- are supposed to be in contact 
with each other, so that their contact potential energy cannot 
be dissipated. 
 
The deviation from the classical inelastic hard sphere HCS is 
$J(E,\lambda_c) = J(\lambda_c) \propto E^{0}$, 
a constant independent of the energy and thus independent of time.
Thorough calculation \cite{goldshtein03} yields
\begin{equation} 
J(\lambda_c)=\exp(-k \varepsilon_\lambda) ~, 
\end{equation} 
with
$\varepsilon_\lambda=\lambda_c (N/V) (4a)^2 g_{2a}(\nu)
                    =\lambda_c / \sqrt{\pi} \lambda$,
and constant $k \approx 7.37$.
This result can be understood, since in the homogeneous cooling 
regime, one has constant density and thus constant mean free 
path, so that a free path cut-off model has a time independent 
effect.  Due to its lack of interesting new phenomena for the  
HCS, we will not discuss the LC model further. \\
 
The {\em TC model} was invented in order to model elastic material properties,
like the ``detachment'' effect \cite{luding94d}, in the framework of the 
IHS model. In soft assemblies of particles this resembles multi-particle 
contacts and avoids the inelastic collapse in dense IHS systems
\cite{luding96e,luding97c,luding98f};   
the physical idea behind was discussed in the introduction.
In technical terms, a collision is elastic if any one of two colliding
particles had a collision within a time $t_c$ before the actual time. 

The deviation from the classical HCS is, see \cite{goldshtein03},  
\begin{equation} 
J(E,t_c) 
         = \exp \left ( \Psi(x) \right ) ~,
\label{eq:Jtc} 
\end{equation} 
with the series expansion
$\Psi(x)=-1.268x+0.01682x^2-0.0005783x^3+{\cal O}(x^4)$
in the collision integral, with 
$x=\sqrt{\pi} t_c t_E^{-1}(0) \sqrt{E}=\sqrt{\pi} \tau_c(0) \sqrt{E}
=\sqrt{\pi} \tau_c$ \cite{goldshtein03}. 
This is close to the result $\Psi_{\rm LM}=-2x/\sqrt{\pi}$, 
proposed by Luding and McNamara, based on probabilistic mean-field 
arguments \cite{luding98f} \footnote{$\Psi_{\rm LM}$ thus neglects 
non-linear terms and underestimates the linear part}
Here, the argument of the exponential is proportional to the collision rate  
$t_E^{-1} \propto \sqrt{E}$, different from the other models, so that
$J \propto \exp(\sqrt{E})$.

\begin{figure}[tb]
\noindent
\includegraphics[width=5.8cm,angle=-90]{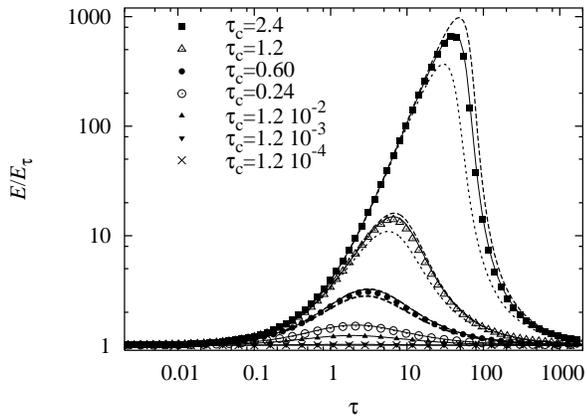}
~\vspace{0.4cm}\\
\caption{Deviation from the HCS, i.e.\ rescaled energy $E/E_\tau$,
plotted against $\tau$ for simulations with different $\tau_c(0)$ as
given in the inset, with $r_0=0.99$, and $N=8000$.  Symbols are simulation
results, the dashed line is the first order correction, the solid
line results from the third order, and the dotted line correspond to the
results from \protect\cite{luding98f}.
}
\label{fig:tc2}
\end{figure}

Simulation results are compared to the theory in Fig.\ \ref{fig:tc2}.
The agreement between simulations and theory is almost perfect in the examined
range of $t_c$-values, only for large deviations from the HCS solution
and for large $t_c$-values, a few percent discrepancy are observed
\footnote{The simulation has to be
carefully prepared for the TC model in order to achieve good agreement.
First, the system is relaxed elastically with $r=1$ for several hundred
collisions per particle and the last time of collision is saved for each
particle.
Second, the dissipation is switched on and the TC model is activated
using the saved information about previous contacts.  If this information
is not used, one observes an artificial initial decay of energy in the 
simulation.}

The results can be explained as follows. The fixed cut-off time
has no effect when the time between collisions is very large $t_E \gg t_c$,
but strongly reduces dissipation when the collisions occur with high
frequency $t_E^{-1} \stackrel{>}{\sim} t_c^{-1}$.
Thus, in the homogeneous cooling state, there is a strong effect initially,
but the long time behavior tends towards the classical decay
$E \propto t^{-2}$.


Additional simulations with a set of system-sizes and for 
different (also very small ) restitution coefficients will be
discussed elsewhere \cite{goldshtein03}.  Note however, that our
conclusions are valid for all system sizes examined and for 
arbitrary restitution coefficients before the inhomogeneities evolve.

In summary, a general class of cut-off models was presented, aiming 
towards the enhancement of classical kinetic theory with respect to the 
realistic behavior of dissipative particles in the presence of multi-particle 
interactions \footnote{Only the TC model was used for the detailed 
discussion}. 
Analytical expressions for the collisional cooling rate in the energy 
balance equation of the hydrodynamic equation is provided for the 
multi-particle contacts, evading the singularity of the inelastic collapse.
Our theoretical results were verified by event-driven numerical simulations
of the HCS and perfect agreement was obtained.
%
%
For realistic
material behavior combinations of the models and also refinements may
be neccessary. Our model, however, leads to a correction of the
energy dissipation term alone, in the framework of a hydrodynamic 
continuum theory. We regard it thus as much simpler than the model
proposed in Ref.\ \cite{hwang95} that also takes the finite contact 
duration into account.

The TC model, and to some extent also the other models, are based on 
the assumption that the elastic, reversible, potential contact energy of 
real particles cannot be dissipated in the same way as the kinetic energy.
If one has multi-particle contacts in the system, a lot of energy 
is stored in their contacts -- and thus cannot be dissipated. 
%
 
Future interesting work could be the extension of our cut-off models 
to more complicated material laws, e.g., introducing some velocity  
dependent restitution coefficient $r(v)$ or contact duration $t_c(v)$. 
In the same spirit, the cut-off law can be replaced by continuous functions
instead of step-functions \footnote{However, since experimental data
are missing, we prefer the simple event-based model which is consistent
with the kinetic theory}.
In addition, the present theory should be applied to hydrodynamic models 
of inhomogeneous systems, where the cut-off criterion is a function of 
the position, in order to prove its general applicability. As another 
verification, the model could be compared to soft-sphere simulations and 
experiments. 




\end{document}